\documentclass[12pt]{article}
\usepackage[T1]{fontenc}
\usepackage[utf8]{inputenc}

\usepackage[backend=biber,
  sorting=none, 
  autocite=plain, 
  url=true,                  
  hyperref=true,
  style=numeric-comp]{biblatex}
\usepackage{graphicx}
\usepackage{subfig}
\usepackage{amsmath}
\usepackage{amssymb}
\usepackage[section]{placeins}
\usepackage{pgfplots}
\pgfplotsset{compat=1.12}
\usepackage{xcolor}
\definecolor{ballblue}{rgb}{0.13, 0.67, 0.8}

\usepackage{bm}
\usepackage{physics}

\usepackage{mathtools}
\usepackage{esvect}

\usepackage{fontawesome}
\usepackage{enumitem}

\usepackage[margin=0.75in]{geometry}

\usepackage[colorlinks=true, citecolor=blue]{hyperref}

\usepackage{authblk}

\usepackage{caption}
\title{Prospects of Indirect Detection for the Heavy S3 Dark Doublet}

\author[1]{C. Espinoza\footnote{m.catalina@fisica.unam.mx}}
\author[2]{M. Mondrag\'{o}n\footnote{myriam@fisica.unam.mx}}
\affil[1]{C\'atedras Conacyt - Instituto de F\'{i}sica, Universidad Nacional Aut\'{o}noma de M\'{e}xico, Apdo. Postal 20-364, Cd. M\'{e}xico 01000, M\'{e}xico.}
\affil[2]{Instituto de F\'{i}sica, Universidad Nacional Aut\'{o}noma de M\'{e}xico, Apdo. Postal 20-364, Cd. M\'{e}xico 01000, M\'{e}xico.}
\date{}                   
\begin{document}
	
\maketitle
	
\begin{abstract}

We present an analysis of the Dark S3 model in the heavy DM mass region, this model
features an S3 symmetric extension of the scalar sector of the SM including a scalar $SU(2)$ doublet 
dark matter candidate. We use publicly available tools to compute, in addition to conventional 
physical constraints on the parameter space of the model, the Sommerfeld enhancement factors
for the present day annihilation cross section and the likelihood profile for a simulation of 
an observation run by the Cherenkov Telescope Array of the Coma Berenices dwarf galaxy. 
Our results disfavour masses above $\sim$5 TeV mainly because of overproduction of dark matter
not consistent with the relic abundance observations; we also find a moderately large region with
masses in between 1.2 and 4.9 TeV which predict the correct value of the DM relic and in addition
have a light scalar with the characteristics of the SM Higgs boson (i.e. within the decoupling
limit akin to the THDM). Comparison of our results with model independent exclusion curves from
HESS and other CTA simulations show that these limits fall short only an order of magnitude
in the value of the annihilation cross section in order to exclude the best fit point of the model.

\end{abstract}

\section{Introduction}
\label{intro}

The study of extensions of the Standard Model (SM) capable of tackling
one or more of the well known issues present in this paradigm of particle physics
continues to be one of the most active fields of contemporary research.
One strategy is to approach the subject from the scalar sector
of the SM, enlarging its field content with extra scalars while keeping the rest
of the sectors untouched. 
As demonstrated by the vast literature on the Two Higgs Doublet Model (THDM),
the simplest of such extensions,  
the possibility of stumbling into rich and interesting
phenomenology is just around the corner (see e.g.~\autocite{Branco:2011iw} and references therein). 
Moreover, by taking the extra scalar doublet
as inert (the Inert 
Doublet Model (IDM)~\autocite{Deshpande:1977rw}) we end up with a very simple and at the same time
highly illustrative model containing a candidate for dark matter
with rich phenomenology, e.g. see for example~\autocite{McDonald:1993ex,Ma:2006km,Barbieri:2006dq,LopezHonorez:2006gr,Majumdar:2006nt,
	Hambye:2007vf,Gustafsson:2007pc,Cao:2007rm,
	Agrawal:2008xz,Lundstrom:2008ai, Hambye:2009pw,Andreas:2009hj,Nezri:2009jd,Arina:2009um,Dolle:2009ft,
	LopezHonorez:2010tb,Honorez:2010re,Ginzburg:2010wa,Hirsch:2010ru,Miao:2010rg,
	Arhrib:2012ia,Swiezewska:2012eh,
	Klasen:2013btp,Garcia-Cely:2013zga,Krawczyk:2013jta,Goudelis:2013uca,
	Garcia-Cely:2015khw,Queiroz:2015utg,
	Belyaev:2016lok,
	Eiteneuer:2017hoh,Biondini:2017ufr,Dutta:2017lny,
	Wan:2018eaz,Belyaev:2018ext,Kalinowski:2018kdn}.

Multihiggs models are natural generalizations of this scheme, including those with
additional symmetries imposed, for example the $S3$ symmetric model where a total of 
three Higgs doublets are present and it is assumed that these scalars (and the matter sector) 
belong to irreducible representations of the permutation group $S3$ 
reflecting a hypothesized discrete symmetry of the model,
motivated in part by the fact that matter particles of a given flavor are
only distinguishable from other flavors by their masses and therefore are identical to each other before electroweak
symmetry breaking (EWSB). The study of these models was pioneered in~\autocite{Pakvasa:1977in,Derman:1978rx,Derman:1979nf}
which has led to subsequent interesting research,
see e.g.~\autocite{Mondragon:1998ir,Kubo:2003iw,Kubo:2004ps,Mondragon:2007af,
	Canales:2012dr,Teshima:2012cg,Canales:2013cga,Das:2014fea, 
	Barradas-Guevara:2014yoa,Emmanuel-Costa:2016vej,Das:2015sca,
	Barradas-Guevara:2015rea,
	Espinoza:2018itz,Gomez-Izquierdo:2018jrx,
	Garces:2018nar,
	Mishra:2019keq,
	Chakrabarty:2019tsm,Kuncinas:2020wrn,CarcamoHernandez:2020pxw}.

In this letter we continue our research initiated in \autocite{Espinoza:2018itz} 
wherein we exploit the fact that the permutation group $S3$ has four irreducible representations,
this leaves just enough space to augment to the $S3$ symmetric model a fourth scalar doublet
which, when taken as inert, comprises a simple IDM-like dark sector.
This time, we focus our attention on the high ($1-20$ TeV) DM candidate mass region of the
parameter space of the model exploring the possibilities of indirect detection (ID) of
gamma ray signals specifically in the Cherenkov Telescope Array (CTA), from a simulated observation
of the dwarf spheroidal galaxy Coma Berenices\footnote{
	As explain in section \ref{numerics} due to the large dimensionality of the parameter
	space the work load of the calculation of the likelihood profile makes analyzing more
	than one dwarf rather unfeasible with the computational resources at our disposal.
}.
In this large mass regime it is important to take into account the nonperturbative phenomenon 
of Sommerfeld enhancement which is a manifestation of the multiple exchanges of gauge bosons and
higgses between two annihilating DM particles, specially for the correct analysis of the gamma
ray fluxes predicted by the model whose detection is the holy grail of dark matter ID experiments. 
This phenomenon occurs because the DM particles move at present with nonrelativistic (NR) velocities
($\sim 10^{-3}$) and (colloquially speaking) they have enough time to exchange (on-shell) mediators
before the actual annihilation. As a result, the wave function of the DM pair is no longer
a plane wave and to correctly predict the annihilation cross section it is necessary to find out
the modified wave function. This process becomes particular important when the mass of the DM
candidate is much larger than the gauge bosons and higgses because then the electroweak force 
between the DM pair becomes ``long range'', 
and also when there are other particles in the dark sector such as those part of the same multiplet of $SU(2)$ as the 
DM candidate, because then the EWSB induced mass splitting between these particles is relatively small
and as a result the exchange of mediators can induce transitions of the original annihilating 
DM pair to other states in the multiplet, which potentially enhances the annihilation cross section.

Thus, for our analysis we compute the predicted DM annihilating gamma ray flux with Sommerfeld corrections
from the mentioned dwarf galaxy and use this prediction to obtain a likelihood estimate from a simulated
observation at the CTA. Specifically, we simulate a null-result experiment, meaning that we assume that
during the observation period no significant gamma ray signal above background was encountered, this permits us to obtain 
a likelihood estimate and a test statistics (TS) from the comparison of both hypothesis, the existence or
non-existence of a signal from DM annihilation, which in turn leads to the estimate of limiting or exclusion
regions for the parameter space of the model. In order to obtain such information of the parameter space
we perform a directed scan, for each probed point (set of values of the independent parameters) in the model 
we feed the predicted flux to the public CTA analysis tools obtaining a likelihood estimate and a TS, we then 
supplement the estimated likelihood with information regarding unitarity and collider constraints on
heavy scalars as well as the comparison with the relic abundance experimental value. For simplicity, we neither reject
points on account of violating unitarity or collider bounds nor define additional likelihood functions
to deal with these constraints. Instead, equivalently, we simply penalize the CTA likelihood estimate when this happens
as well as when the predicted relic abundance is above the experimental value, but we allow for points
with relic abundance predictions below this bound to account for the possibility of under-abundant DM component.
Our analysis allows us to present a likelihood profile of the parameter space of the model and estimate to what
extend future observations of the CTA array can probe the model.

\section{Flux from annihilating DM pairs}

\subsection{The model}

The $S3$-model is an extension of the scalar sector of the SM with a total of three $SU(2)$ doublets
which, together with the matter sector, are assumed to transform under the permutation group $S3$
in such a manner that the Lagrangian respects the symmetry even after EWSB. Two of the EW doublets,
$\Phi_1$ and $\Phi_2$ are chosen to form an $S3$ doublet while the third one $\Phi_s$ transforms 
as a symmetric singlet of $S3$; the matter sector is chosen to have transformation properties
as in reference \autocite{Canales:2013cga}. We take advantage of the fact that the $S3$ group has
four irreducible representations which allows us to introduce a dark sector in the model by
including an extra $SU(2)$ doublet, $\Phi_a$, transforming as an antisymmetric singlet of $S3$ and imposing
a $Z_2$ discrete symmetry under which the only field with nontrivial transformation is $\Phi_a$.
The dark sector scalar potential is constructed as follows:

\begin{equation}
V_{\textrm{DM}} = V_{2a} + V_{4a} + V_{4sa}
\end{equation}
where each term is given respectively by:

\begin{equation}
V_{2a} = \mu_2^2 \Phi_a^\dagger \Phi_a
\end{equation}

\begin{eqnarray}
\label{pot4a}
V_{4a} & = & \lambda_{10}(\Phi_{a}^{\dagger}\Phi_{a})(\Phi_{1}^{\dagger}\Phi_{1}+\Phi_{2}^{\dagger}\Phi_{2}) \nonumber
\\ &  & +\lambda_{11}[(\Phi_{a}^{\dagger}\Phi_{1})(\Phi_{1}^{\dagger}\Phi_{a})+(\Phi_{a}^{\dagger}\Phi_{2})(\Phi_{2}^{\dagger}\Phi_{a})] \nonumber
\\ &  & +\lambda_{12}[(\Phi_{a}^{\dagger}\Phi_{1})(\Phi_{a}^{\dagger}\Phi_{1})+(\Phi^{\dagger}_{a}\Phi_{2})(\Phi^{\dagger}_{a}\Phi_{2})+\mathrm{h.c.}] \nonumber
\\ &  & +\lambda_{13}(\Phi^{\dagger}_{a}\Phi_{a})^{2} 
\end{eqnarray}

\begin{equation}
V_{4sa} = \lambda_{14} (\Phi_{s}^{\dagger} \Phi_{a}   \Phi_{a}^{\dagger} \Phi_{s})
\end{equation}
For the full Lagrangian and complete details of the model we refer the reader to our previous work \autocite{Espinoza:2018itz},
here we'll just give some expressions not explicitly given there which we'll use in the recount of this research.
After EWSB the $Z_2$ even scalars acquire vacuum expectation values (vev) $v_1$, $v_2$ and $v_s$, 
but from the consistency of the minimization conditions of the scalar potential two of the vevs are aligned
$v_1=\sqrt{3}v_2$. This, together with the requirement $v=\sqrt{v_s^2 + 4 v_2^2}=$ 246 GeV, where $v$ is the SM vev
leads to only one independent vev parameter, we choose $\tan{\theta} = 2v_2 / v_s$ as independent parameter in the
numerical calculations. EWSB induces mixing between the $Z_2$ even scalars, the CP even ones mix through the matrix:

\begin{equation}\label{rotM}
Z^h =
\left(
\begin{array}{ccc}
 \cos(\alpha) & 0 & \sin(\alpha) \\
 0 & 1 & 0 \\
 -\sin(\alpha) & 0 & \cos(\alpha) \\
\end{array}
\right)
\left(
\begin{array}{ccc}
 1 & 0 & 0 \\
 0 & \frac{1}{2} & -\frac{\sqrt{3}}{2} \\
 0 & \frac{\sqrt{3}}{2} & \frac{1}{2} \\
\end{array}
\right)
\end{equation}
while the charged and CP-odd scalars mix respectively with the matrices:

\begin{equation}\label{rotM2}
Z^C = Z^A =
\left(
\begin{array}{ccc}
 \cos ({\theta}) \sin ({\theta}) & 0 & \sin ^2({\theta}) \\
 0 & 1 & 0 \\
 -\cos ({\theta}) \sin ({\theta}) & 0 & \cos ^2({\theta}) \\
\end{array}
\right)
\left(
\begin{array}{ccc}
	1 & 0                  & 0 \\
	0 & \frac{1}{2}        & -\frac{\sqrt{3}}{2} \\
	0 & \frac{\sqrt{3}}{2} & \frac{1}{2} \\
\end{array}
\right)
\end{equation}
i.e. both matrices coincide.
Explicit expressions for the mixing angle $\alpha$ and the masses of all the scalars
can be found in \autocite{Espinoza:2018itz}, these are given in terms of the 15 free
parameters of the model: $\lambda_1$ - $\lambda_8$, $\lambda_{10}$ - $\lambda_{14}$, $\tan{\theta}$ and $\mu_2$
(there are only 15 free parameters because we are assuming a simple form of the Yukawa matrices,
see the above reference for details and justification).
There are 10 physical scalars in the model, we use interchangeably the following two notations,
$\tilde{h}_k$ with $k=1,2,3$ for the neutral scalars $H$, $H_3$ and $h$ 
(i.e. $\tilde{h}_1 = H$, $\tilde{h}_2=H_3$ and $\tilde{h}_3=h$), 
$\tilde{A}_k$ for the neutral pseudo-scalars $G^0$, $A_3$ and $A$, and
$\tilde{H}^\pm_k$ for the charged scalars $G^\pm$, $H_3^\pm$ and $H^\pm$, 
and in the dark sector we have the fields $h_a$, $A_a$ and $H_a^\pm$ with $h_a$ the DM candidate. 
Here $G^0$ and $G^\pm$ are the
Goldstone fields which must be taken into account when it is advantageous to work in Feynman gauge.
The second notation is intended to facilitate comparison with the Two Higgs Doublet Model (THDM)
since a subset of the field content of the $S3$ model resembles the corresponding content of the THDM.
It is favorable to work with the set of physical masses of the scalars as independent variables
alongside with the parameters $\mu_2$, $\lambda_{13}$, $\lambda_{14}$, $\tan{\theta}$ and $\alpha$,
so we invert the expressions for the masses obtaining:

\begin{equation}\label{lamdas}
\begin{array}{l@{}l}
\lambda_1      &{}= \left(\csc^2\theta (9 \cos (2 {\alpha}) ({M^2_h}-{M^2_H})+9 {M^2_h}+9 {M^2_H}+18 {M^2_{H_3^\pm}}-2 {M^2_{H_3}}-18 {M^2_{H^\pm}})+18 {M^2_{H^\pm}} \right) /(36 v^2) \\
\lambda_2      &{}= \left(\csc^2\theta (-{M^2_{A_3}}+{M^2_{H_3^\pm}}+{M^2_A}-{M^2_{H^\pm}})-{M^2_A}+{M^2_{H^\pm}}\right) / (2 v^2)   \\ 
\lambda_3      &{}= \csc^2\theta (9 {M^2_{H^\pm}} \cos (2 {\theta})-18 {M^2_{H_3^\pm}}+8 {M^2_{H_3}}+9 {M^2_{H^\pm}}) / (36 v^2)  \\
\lambda_4      &{}= -2 M^2_{H_3} \csc\theta \sec\theta / (9 v^2)  \\ 
\lambda_5      &{}= \left( 9 \sin (2 {\alpha}) \csc\theta \sec\theta ({M^2_H}-{M^2_h})+2 {M^2_{H_3}} \sec^2\theta+36 {M^2_{H^\pm}} \right) / (18 v^2)  \\
\lambda_6      &{}= \left( {M^2_{H_3}} \sec^2\theta+9 {M^2_A}-18 {M^2_{H^\pm}}\right) / (9 v^2)   \\
\lambda_7      &{}= \left( {M^2_{H_3}} \sec^2\theta-9 {M^2_A} \right) / (18 v^2)  \\
\lambda_8      &{}= \sec^2\theta \left(9 \cos (2 {\alpha}) ({M^2_H}-{M^2_h})-2 {M^2_{H_3}} \tan ^2({\theta})+9 ({M^2_h}+{M^2_H})\right) / (36 v^2) \\
\lambda_{10}   &{}= 2 ({M^2_{H_a^\pm}}-{\mu^2_2}) \csc^2\theta / v^2  \\
\lambda_{11}   &{}= -\left( {\lambda_{14}} v^2 \cot\theta^2-{M^2_{h_a}} \csc^2\theta-{M^2_{A_a}} \csc^2\theta+2 {M^2_{H_a^\pm}} \csc^2\theta \right) / v^2 \\
\lambda_{12}   &{}= \left( \csc^2\theta (M^2_{h_a} - M^2_{A_a}) \right) / (2 v^2)
\end{array}
\end{equation}
These expressions are useful because in the directed scan it is simpler to vary the masses to 
probe the regions we are interested the most.

\subsection{Enhanced annihilating cross section}

As explained in the introduction, in the region of large DM masses it is important to take
into account the Sommerfeld enhancement of the annihilation cross section, specially since we
will assume the DM sector quasi-degenerate in mass $M_{h_a} \simeq M_{A_a} \simeq M_{H_a^\pm}$. These type of corrections
are mandatory to include since the perturbative calculation of this cross section violates unitarity
bounds in the large DM mass limit already at the one loop order, as can be seen for example in the
process of DM annihilation into a pair of gammas which does not happens at the tree level.
Thus, higher order corrections are needed to predict correctly this observable. The dominant contributions
of higher order are the ladder type diagrams where multiple exchanges of on-shell mediators take place
before the actual annihilation of the DM pairs. A complete formulation for the treatment of
Sommerfeld corrections can be found e.g. in~\autocite{Hisano:2002fk,Hisano:2004ds,Hisano:2003ec};
for our model we follow their handling as done for the IDM in~\autocite{Garcia-Cely:2015khw} since both dark sectors
are very much alike. 

As a first step it is necessary to find out the effective NR potentials
induced by the exchange of the different gauge bosons and scalars, this is most simply done by calculating
the allowed tree level scattering amplitudes in the complete theory and directly take the NR
limit on them; also importantly to remember is to rescale the fields in order to have the NR
Hamiltonian in canonical form otherwise the correct dependence of the potentials on the DM mass won't be attained.
Also, it is useful to display the potential as a matrix in a basis of two-particle states in accordance
with the possible scattering amplitudes $h_a h_a \rightarrow \mathrm{DM DM}$, $A_a A_a \rightarrow \mathrm{DM DM}$
and $H_a^+ H_a^- \rightarrow \mathrm{DM DM}$ where $\mathrm{DM DM}$ here means either of $h_a h_a$, $A_a A_a$ or 
$H_a^+ H_a^-$. Note that the first process here, on account of the mass quasi-degeneracy,
can induce the other two as part of the multi-scattering ``ladder''
but those latter processes cannot happen on its own reflecting the fact that $h_a$, being the DM candidate,
is the only DM particle of the model present today. Explicitly the basis of two-particle states is
$\{h_a h_a, A_a A_a, H_a^+ H_a^-\}$ and the elements of the matrix potential in this basis are obtained
from the Fourier transform of the NR amplitude, for instance:

\begin{equation}
V_{1 1}(r) = \frac{1}{4M^2_{h_a}} \int \frac{d^3 q}{(2\pi)^3} e^{i \mathbf{q}\cdot\mathbf{r}} 
i A^{\textrm{2-body}}_{\textrm{NR}}(h_a h_a \rightarrow h_a h_a)
\end{equation}
where $r$ is the relative coordinate of the two-particle state; the prefactor $1/4M^2_{h_a}$
comes from the rescaling of the NR fields. In general, the two body amplitudes in momentum 
space $A^{\textrm{2-body}}_{\textrm{NR}}$ differ by a factor of $\sqrt{2}$ relative to the ``raw'' amplitude
calculated from the tree level Feynman diagram due to symmetrization of the two particle state
when it is form by identical particles in the initial state and distinguishable particles in the
final state or vice versa. To understand this factors it is useful to visualize this amplitudes
in the equivalent basis 
$\{h_a h_a, A_a A_a, H_a^+ H_a^-, H_a^- H_a^+ \}$
where the last two states
are treated as different; the equivalency of both basis is discussed in~\autocite{Beneke:2014gja}
in the context of the MSSM.

In this manner we obtain for the potential matrix:

\begin{equation}\label{pot}
V(r) = V_1(r) + V_2(r)
\end{equation}

\begin{equation}\label{pot1}
V_1(r) =
\left(
\begin{array}{ccc}
 \frac{{ -|s_{\tilde{h}_k}^{h_ah_a}|^2 e^{-M_{\tilde{h}_k} r}}}{16\pi M_{h_a}^2 r}  
 & 
 \frac{-{ |s_{\tilde{A}_k}^{h_aA_a}|^2 e^{-M_{\tilde{A}_k} r}}}{16\pi M_{h_a}^2 r}  
 & 
 \frac{-{ |s_{\tilde{H}^\pm_k}^{h_aH_a^\pm}|^2 e^{-M_{\tilde{H}^\pm_k} r}}}{16\pi M_{h_a}^2 r}  
 \\
 \frac{-{ |s_{\tilde{A}_k}^{h_aA_a}|^2 e^{-M_{\tilde{A}_k} r}}}{16\pi M_{h_a}^2 r}  
 & 
 \frac{-{ |s_{\tilde{h}_k}^{A_aA_a}|^2 e^{-M_{\tilde{h}_k} r}}}{16\pi M_{h_a}^2 r}  
  {\scriptstyle + 2 (M_{A_a} - M_{h_a})}
 & 
 \frac{-{ |s_{\tilde{H}^\pm_k}^{A_aH_a^\pm}|^2 e^{-M_{\tilde{H}^\pm_k} r}}}{16\pi M_{h_a}^2 r}  
 \\
 \frac{-{ |s_{\tilde{H}^\pm_k}^{h_aH_a^\pm}|^2 e^{-M_{\tilde{H}^\pm_k} r}}}{16\pi M_{h_a}^2 r}  
 & 
 \frac{-{ |s_{\tilde{H}^\pm_k}^{A_aH_a^\pm}|^2 e^{-M_{\tilde{H}^\pm_k} r}}}{16\pi M_{h_a}^2 r}  
 & 
 \frac{-{ |s_{\tilde{h}_k}^{H_a^\pm H_a^\pm}|^2 e^{-M_{\tilde{h}_k} r}}}{16\pi M_{h_a}^2 r} 
   {\scriptstyle + 2 (M_{H^\pm_a} - M_{h_a})}
 \\
\end{array}
\right)
\end{equation}

\begin{equation}\label{pot2}
V_2(r) =
\left(
\begin{array}{ccc}
 0
 & 
 -\frac{g_2^2 e^{-M_Z r} }{16\pi c_w^2 r} 
 & 
 -\frac{g_2^2 e^{-M_W r}}{16\pi r}  
 \\
 -\frac{g_2^2 e^{-M_Z r} }{16\pi c_w^2 r}  
 & 
 0
 & 
 -\frac{g_2^2 e^{-M_W r}}{16\pi r}  
 \\
 -\frac{g_2^2 e^{-M_W r}}{16\pi r}  
 & 
 -\frac{g_2^2 e^{-M_W r}}{16\pi r} 
 & 
 -\frac{g_2^2 (s_w^2 + (1 - 2 s_w^2)^2 e^{-M_Z r})}{16\pi c_w^2 r}
 \\
\end{array}
\right)
\end{equation}
with implicit sums over $k$ and we have included the terms corresponding to the mass splittings; in the numerical calculation
we will vary the two mass gaps as free parameters instead of the masses $M_{A_a}$, and $M_{H_a^\pm}$.
Here $g_2$ is the weak coupling constant, $s_w=\sin\theta_w$, $c_w=\cos\theta_w$ with
$\theta_w$ the Weinberg angle and the couplings are of the form:

\begin{equation}\label{coups}
\begin{array}{l@{}l}
s_{\tilde{h}_k}^{h_ah_a}           &{}= -i [v\sin\theta (\lambda_{10} + \lambda_{11} + \lambda_{12})
                                   (\sqrt{3}Z^h_{k2}+Z^h_{k3})/2 + v\cos\theta \lambda_{14} Z^h_{k1}] \\
s_{\tilde{h}_k}^{A_aA_a}           &{}= -i [v\sin\theta (\lambda_{10}+\lambda_{11} - 2 \lambda_{12})
                                   (\sqrt{3}Z^h_{k2}+Z^h_{k3})/2 + v\cos\theta \lambda_{14} Z^h_{k1}] \\ 
s_{\tilde{h}_k}^{H_a^\pm H_a^\pm}  &{}= -i\lambda_{10} v\sin\theta (\sqrt{3}Z^h_{k2}+Z^h_{k3})/2 \\
s_{\tilde{A}_k}^{h_aA_a}           &{}= -2i\lambda_{12} v\sin\theta(\sqrt{3}Z^A_{k2}+Z^A_{k3})/2 \\ 
s_{\tilde{H}^\pm_k}^{h_aH_a^\pm}   &{}= (-i/2)[\lambda_{14} v\cos\theta Z^C_{k1} + 
                                        v\sin\theta(2\lambda_{12}+\lambda_{11})(\sqrt{3}Z^C_{k2}+Z^C_{k3})/2] \\ 
s_{\tilde{H}^\pm_k}^{A_aH_a^\pm}   &{}= (-i/2)[\lambda_{14} v\cos\theta Z^C_{k1} + 
                                        v\sin\theta(-2\lambda_{12}+\lambda_{11})(\sqrt{3}Z^C_{k2}+Z^C_{k3})/2] \\ 
\end{array}
\end{equation}

\begin{figure}[t]%
    \centering
    \subfloat[]{{\includegraphics[width=0.5\textwidth]{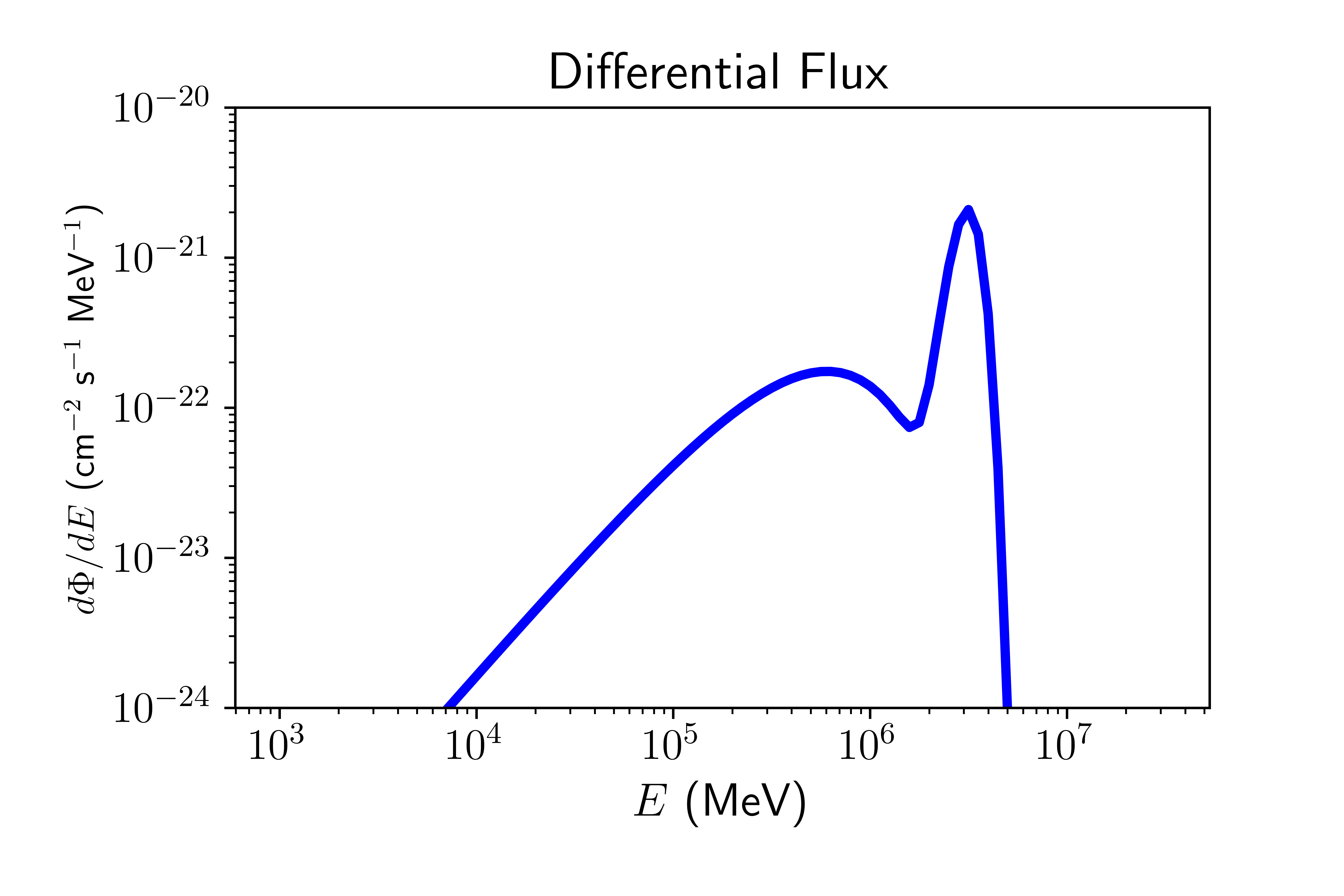} }}%
    \subfloat[]{{\includegraphics[width=0.5\textwidth]{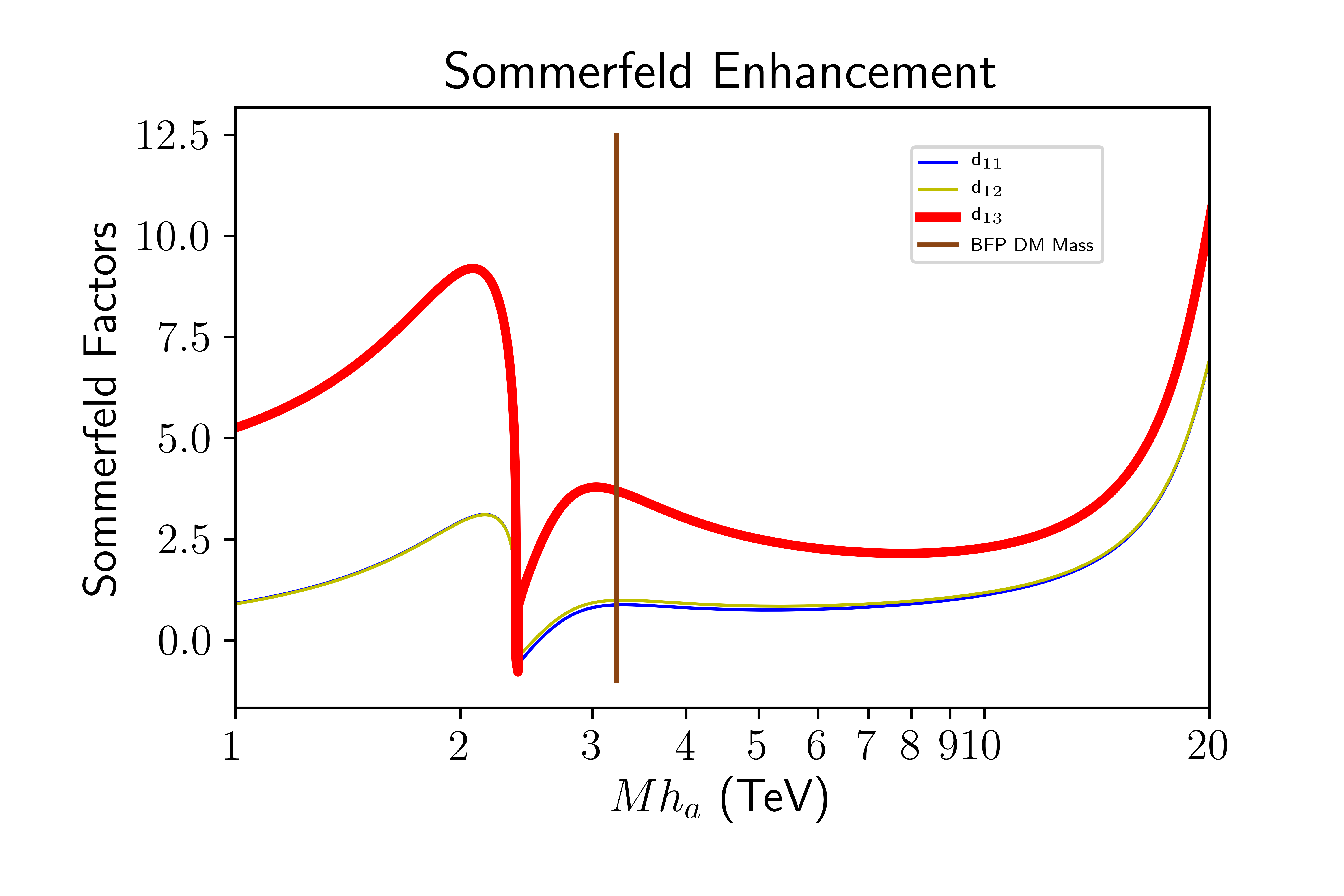} }}%
    \caption{(a) Differential flux from DM pair annihilation (eq. \ref{flux}) for the benchmark point of the model computed in section \ref{results}.
    (b) Sommerfeld enhancing factors as a function of the DM candidate mass, other parameters are taken as in
    the benchmark point. The vertical line signals the value of the benchmark DM mass.}%
	\label{fig:somm1}%
\end{figure}

With the matrix potential at hand, a Schr\"{o}dinger like system of equations for the deformed wave function~\cite{Hisano:2004ds}
is solve numerically. Here we follow~\autocite{Garcia-Cely:2015dda} where an equivalent equation with advantageous
numerical properties is put forward:

\begin{equation}\label{Sch}
h^\prime(r) + h^2(r) + \frac{1}{4} M^2_{h_a} v_{\textrm{rel}}^2 - M_{h_a} V(r) = 0
\end{equation}
with the matrix $h(r)$ satisfying the boundary condition

\begin{equation}
h(\infty) = (i M_{h_a} v_{\textrm{rel}} / 2)\sqrt{1 - 4V(\infty) / (M_{h_a} v_{\textrm{rel}}^2)}
\end{equation}
where $v_{\textrm{rel}} \sim 2\times 10^3$ is the present day relative velocity of annihilating DM.
The column vector defined as $d\equiv(d_{11} \, d_{12} \, d_{13})^\textrm{T}$ contains the Sommerfeld
factors which are calculated from the relation:

\begin{equation}\label{Sfactors}
d d^\dagger = \frac{1}{i M_{h_a} v_{\textrm{rel}}} (h(0) - h^\dagger(0))
\end{equation}
the matrix $d d^\dagger$ has only one nonzero eigenvalue corresponding to its eigenvector $d$, note that
this eigenvalue is just the modulus square of $d$. 
In figure \ref{fig:somm1} (b) we show the variation of the
Sommerfeld factors with the DM mass, other free parameters are taken as in the benchmark or Best Fit Point (BFP) of section 
\ref{results}. We note that the three factors become slightly negative and close to zero around a DM mass of $\sim$ 2.3 TeV,
suggesting that around these values of parameter space a marked destructive interference is at work.
On the other hand, for masses below 5 TeV the maximum occurs around $\sim$ 2.1 TeV, with another local maximum
close to $\sim$ 3 TeV, with enhancements not surpassing a factor of ten. The BFP does not occurs in the vicinity
of the highest enhancements and thus other important physical constraints weight in to shift it to a higher mass.
In figure \ref{fig:somm1} (a) we also present the differential flux for the BFP as computed in the next section, 
including the Sommerfeld corrections.


The enhanced cross sections are then obtained with the aid of the Optical Theorem which relates the Forward Scattering
Amplitude with the (tree level) annihilating amplitudes of the DM particles in the multiplet (for full details see the above
references). The annihilating cross section in the limit of zero relative velocity for a given final state $f$ is given by:

\begin{equation}
\sigma v_{\mathrm{rel}} (h_a \, h_a \rightarrow f) = \frac{1}{2} (D \Gamma^f D^\dagger)_{1 1} 
\end{equation}
where $\Gamma = \sum_f \Gamma^f$ is the total matrix of absorptive terms to all final states $f$, and 
$D$ is the matrix whose only nonzero row (in our basis the first row) is $d^\mathrm{T}$. The elements of
$\Gamma^f$ are given explicitly by

\begin{equation}
\Gamma^f_{i j} = \frac{N_i N_j}{4M^2_{h_a}} \int \prod_{a\in f}\frac{d^3q_a}{(2\pi)^3 2E_a} 
                 {\cal M}(i \rightarrow f) {\cal M}^*(j \rightarrow f) (2\pi)^4 \delta^4(p_i-p_j)
\end{equation}
here the indexes $i$ and $j$ refer to any element of the two-particle basis (the annihilating pair),
$f$ is any allowed final state of non-DM particles in the model, $p_i$ and $p_j$ are the 4-momenta
of the initial and final states and the symmetry factors are given by $N_{h_a h_a} = N_{A_a A_a} = 1/\sqrt{2}$
and  $N_{H_a^+ H_a^-} = 1$, with ${\cal M}$ the tree level corresponding annihilating amplitudes.

We compute $\Gamma^f$ for the following final states: $\gamma \gamma$, $\gamma Z$,
$ZZ$, $W^+W^-$, $HH$, $H_3 H_3$, $hh$ and $Hh$ (for $HH_3$ and $H_3 h$, $\Gamma^f$ is zero). 
For example, $\Gamma^{H_3 H_3}$ is given by:

\begin{equation}\label{H3H3}
\Gamma^{H_3 H_3} =
\frac{1}{32 \pi M_{h_a}^2}
\left(
\begin{array}{ccc}
 \frac{1}{2} \lambda_+^2
 & 
 \frac{9}{2} \lambda_- \lambda_+
 & 
 \frac{9}{\sqrt{2}} \lambda_{10} \lambda_+
 \\
 \frac{9}{2} \lambda_- \lambda_+
 & 
 \frac{1}{2} \lambda_-^2
 & 
 \frac{9}{\sqrt{2}} \lambda_{10} \lambda_-
 \\
 \frac{9}{\sqrt{2}} \lambda_{10} \lambda_+
 & 
 \frac{9}{\sqrt{2}} \lambda_{10} \lambda_- 
 & 
 \lambda_{10}^2 
 \\
\end{array}
\right)
\end{equation}
where $\lambda_\pm \equiv \lambda_{10}+\lambda_{11}\pm 2 \lambda_{12}$.
Note that for this calculations we approximate the DM mass splittings as zero and we also
neglect terms of the form $M_X/M_{h_a}$ with $X$ any gauge boson or scalar, for the numerical 
calculation we thus will keep the masses of the scalars to relatively light values. We list in appendix \ref{mats} the rest
of the matrices.

\subsection{Flux}

The total differential cross section into gammas is given by

\begin{equation}
\frac{d\sigma v_{\mathrm{rel}}}{dE_\gamma} = \sum_f \sigma v_{\mathrm{rel}} (h_a h_a \rightarrow f)
    \times  \frac{dN^f}{dE\gamma}
\end{equation}
Following~\autocite{Garcia-Cely:2015khw}, for the case of continuous yields ($f$ = EW or scalar boson pair as final state)
we use the parametrization~\autocite{Bergstrom:1997fj}:
      
\begin{equation}
\frac{dN^f}{dE\gamma} = (0.73/M_{h_a}) \, x^{1.5} \, \exp{-7.8 \, x}
\end{equation}
with $x = E\gamma / M_{h_a}$.
For the $\gamma \gamma$ or the $\gamma Z$ final states the yield is a Dirac delta
centered respectively at
$M_{h_a}$   or   $M_{h_a} - M_Z^2 / (4M_{h_a})$.
We model the delta as a Gaussian centered at the corresponding energy
and of width equal to (conservatively) 15\% the energy of the line, this
since a delta would be a ``monochromatic line'' which in the context of
ID experiments refers to spectral features with energy width much smaller 
than the energy resolution of the detector, typically 15\% is achieved e.g.
in HESS and therefore such value would be conservative for the CTA.
Thus, with Gaussian width = 0.15 $M_{h_a}$:
$(dN/dE\gamma)^{\gamma \gamma} = 2  \delta(E_\gamma - M_{h_a})
                               = 2  (2.66 / M_{h_a})  \exp( - 22.22  ( x - M_{h_a} )^2 / M_{h_a}^2 )$
and similar for $\gamma Z$.

Finally, the predicted gamma ray flux from annihilation of DM particles
is given by the expression:

\begin{equation}
\label{flux}
\frac{d\Phi_\gamma}{dE_\gamma} = \frac{1}{4\pi} \left( \int_{\Delta \Omega} d\Omega \int_{l.o.s} ds \rho^2_\chi \right)  
                                 \left[ \frac{1}{2M_{h_a}^2} 
                                 \sum_{f} \sigma v_{\mathrm{rel}} (h_a h_a \rightarrow f) \frac{dN^f}{dE_\gamma} \right]
\end{equation}
here we have deliberately separated the equation into two parts, since the computation involved for each part is of very different nature.
The first part (the astrophysical part) is well establish for example in the case of dwarf spheroidal galaxies from astrophysical observations and it is known as the ``J-factor'', we will take the dwarf as a point source with constant J-factor of
$\log_{10} J = 19.52$ (with $J$ in GeV$^2$ cm$^{-5}$)~\autocite{Geringer-Sameth:2014yza}. With the analysis of the previous
section and the expressions for the yields given above we complete the second part and the flux prediction from the model.

\section{Likelihood estimate}

\begin{figure}[t]
\centering
	\includegraphics[width=0.5\textwidth]{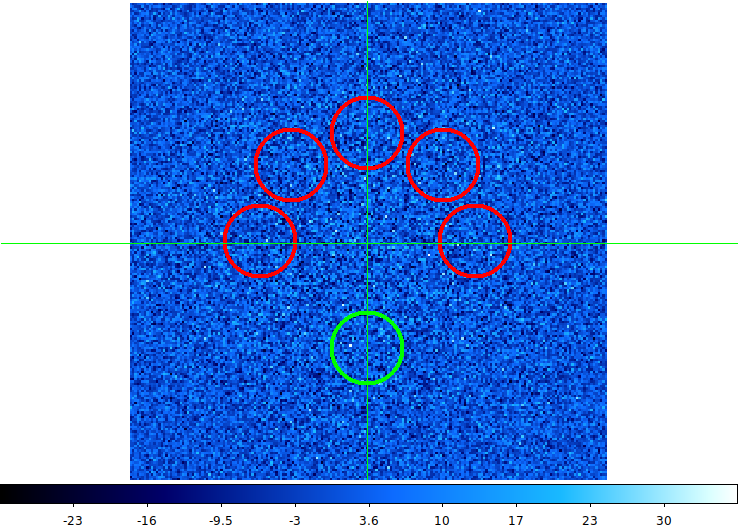}
	\caption{On and Off-regions for the observation of the Coma Berenices spherical dwarf and the determination of the 
	background. The dwarf is located in the center of the green circle at the bottom, and five adjacent regions (red circles) 
	located in symmetric points with respect to the observation point are use
	to determine the background signal, the telescope aim is located at the center of the intersecting lines.
	The color axis represents the signal detection significance, the absence of notorious bright spots in the region means that
	no known sources of gamma rays need to be taken into account in the determination of the background.}
	\label{fig:coma3}%
\end{figure}

We choose to make our analysis based on a simulation of future observations
of the dwarf spheroidal (dSph) galaxy Coma Berenices (or Coma) by the Cherenkov Telescope
Array (southern hemisphere branch). Coma Berenices was discovered in 2006
by the Sloan Digital Sky Survey~\autocite{Belokurov:2006ph}
and is a faint Milky Way satellite at a distance of 44 kpc from the Sun
with right ascension of 12 h 26 min 59 sec and declination of 23 deg 54 min 15 sec;
it is estimated to have a $J$-factor of $\log_{10} J = 19.52$, 
$J$ in GeV$^2$ cm$^{-5}$~\autocite{Geringer-Sameth:2014yza}.
dSph galaxies are recurrent targets for searches of DM signals in ID experiments
by virtue of their large inferred DM density and no known natural sources 
of gamma rays, Coma for example was part of a recent
study of several dSphs in a DM signal search from HESS~\autocite{Abdalla:2018mve};
here, we try to follow their general analysis strategy in as much as possible as 
the available public tools allow us.

We simulate an observation run of Coma of 20 hours with the southern hemisphere CTA array
in which we further assume a no-result experiment, in other words we assume that no significant
excess of gamma rays above nominal background is found throughout the observation period. 
This will allow us to predict, based on current estimations of CTA's expected baseline performance,
exclusion limits on the parameter space of this particular model.
We perform our analysis employing the reflected background technique~\autocite{Aharonian:2006pe}
where the position of the telescope aim is slightly offset from the objective, the latter
defined as a circular region (the On region) centered around Coma from which the simulated observation
is used to fit against the predicted flux from the model. To compute the background in this
technique several twin regions (the Off regions) are defined in symmetrical positions around the telescope
aim (see figure \ref{fig:coma3}) from which the background rate is extracted from the public
Instrument Response Function (IRF) provided by the CTA Consortium. Background rates are expected
to be fairly symmetric with respect to the position of the camera aim and thus this method
minimizes a possible source of systematic errors from modeling the background from Monte Carlo simulations.

In an On/Off analysis the model $M$ is composed of the signal (predicted number of gamma rays)
and the background $M = M_s + M_b$, the background model $M_b$ is taken from the background information 
provided in the IRF and the signal $M_s$ is calculated from the predicted differential gamma ray flux.
The likelihood estimate is constructed from the formula:

\begin{equation}\label{likeliF}
- \log {\cal L}(M) = \sum_i{
	s_i(M_s) + a_i(M_b) b_i(M_b) - 
	n_i^{\mathrm{on}} \log[s_i(M_s) + a_i(M_b) b_i(M_b)] +
	b_i(M_b) - n_i^{\mathrm{off}} \log{b_i(M_b)}
}
\end{equation}
where $n_i^{\mathrm{on}}$ ($n_i^{\mathrm{off}}$) is the number of events in bin $i$ of the On (Off) region,
$s_i(M_s)$ ($b_i(M_b)$) is the number of expected signal (background) counts in bin $i$ of the On (Off) region
and $a_i(M_b)$ is the ratio between the spatial integral over the background model in the On region and
the Off region for bin $i$.

The detection significance of the model is estimated using the Test Statistic (TS) which is defined as:

\begin{equation}\label{TSF}
\mathrm{TS} = 2[\log{{\cal L}(M_s + M_b)} - \log{{\cal L}(M_b)}]
\end{equation}
which involves the log-likelihood value obtained when fitting the model and the background together
to the simulated data, and also the log-likelihood value when fitting only the background.

Note that for the calculation of this quantities all the parameters of the theoretical model remain fixed
because the construction of the likelihood profile is done by an external (to the CTA tools) minimizer
which in each step provides a set of model parameter values along with the predicted differential flux, 
requests the value of the CTA likelihood function (from the CTA tools), 
computes additional likelihood components and based on this information moves around in parameter space searching
for the maximum of the total likelihood function, as explained in the next section.

\section{Numerical analysis}\label{numerics}

The setup for the numerical computation is very simple, we define a function $F$
which accepts as input a set of values for the free parameters of the model
and returns the value of the total likelihood function. The problem is then reduced
to the maximization of this function, for this we use \texttt{Diver}~\autocite{Workgroup:2017htr}
in standalone mode, the differential evolution scanner available in the \texttt{Gambit}~\autocite{Athron:2017ard}
package, and for post-processing of the resulting data sets we use \texttt{Pippi}~\autocite{Scott:2012qh}.
To reduce the number of free parameters of the model in the scan we fix the value of the
scalar $h$ to the Higgs mass~\cite{Aad:2012tfa,Chatrchyan:2012xdj}, for simplicity we fix $\lambda_{13}=0.001$
to easily satisfy inequalities that ensure no tunneling to vacua that breaks the $Z_2$ discrete symmetry~\cite{Ginzburg:2010wa},
and we take $\mu_2^2$ given in terms of the DM mass ($M_{h_a}^2 - \mu_2^2$ small constant) so that
the gaps $M_{H_a^\pm} - M_{h_a}$ and $M_{A_a} - M_{h_a}$ can be taken relatively small
in order to explore regions that presumably have big enhancement factors of the annihilation 
cross section. For the same reason, we take the masses of the heavy scalars in-between 200
and 800 GeV, since the larger the masses of these particles the minor the contribution
to the deforming potential in the Sommerfeld factor.

The layout of $F$ is a little bit more technically challenging though.
Basically we divide points of the model's parameter space into two categories,
the ones that satiate the physical constraints and the ones that doesn't.
In the latter case we don't discard the points, instead equivalently, we assign
a bad likelihood to them. Construction of the constraint filters is done
with the aid of several public tools: we implement the model in 
\texttt{SARAH}~\autocite{Staub:2013tta,Staub:2009bi,Staub:2010jh,Staub:2012pb}
from which we generate corresponding model files for the rest of the tools.
Tree level LQT~\autocite{Lee:1977eg} and finite energy~\autocite{Goodsell:2018tti} 
unitarity conditions are calculated with the 
\texttt{SARAH-\allowbreak SPheno}~\autocite{Staub:2015kfa,Porod:2003um,Porod:2011nf} framework,
and we check current experimental limits from Higgs and heavy scalar searches using 
\texttt{HiggsBounds}~\autocite{arXiv:0811.4169,arXiv:1102.1898,arXiv:1301.2345,arXiv:1311.0055,arXiv:1507.06706}.
For a more complete description of the implementation of the constraints see our previous work
\autocite{Espinoza:2018itz}.

Next we compute the Sommerfeld factors (\ref{Sfactors}) by solving the corresponding
coupled equations (\ref{Sch}) by means of a fourth-order Rosenbrock method~\autocite{10.5555/148286},
the non-relativistic potentials, scattering and annihilation amplitudes are obtained with the aid
of \texttt{FeynArts}~\autocite{Hahn:2000kx}, \texttt{FormCalc}~\autocite{Hahn:1998yk}
and \texttt{FeynCalc}~\autocite{Mertig:1990an,Shtabovenko:2016sxi,Shtabovenko:2020gxv}.
This allow us to calculate the enhanced annihilation cross sections from which we
compute the differential gamma ray flux using (\ref{flux}).

From the predicted differential flux a file with energy and differential flux columns
is created which is then fed to \texttt{ctools}~\autocite{Knodlseder:2016nnv,ctools}, the public
software package for the scientific analysis of Cherenkov Telescope Array data
and simulations. With the aid of \texttt{ctools}, an On/Off analysis of the model is performed
along the lines described in the previous section deriving one piece of the likelihood value
up to this point. We treat Coma as a point-like source and the size of the on and off regions
are taken to be 0.3 degrees in radius. We note that this part of the calculation is the most
expensive in terms of computational time, with so much
free parameters and taking into account that a total of approximately $4.5\times 10^{7}$ $F$-function evaluations
are needed for the algorithm to converge and find the absolute maximum of the likelihood function,
this makes the inclusion of more than one dwarf in the analysis unpractical with
the computational resources at hand,
\footnote{In addition, several runs incrementing the population of the minimizer are necessary to
ensure that proper convergence to the global maximum is attained \autocite{Workgroup:2017htr}.}
 because the position of the dwarfs in the sky are very
different and an independent analysis for each one of them has to be done and therefore the
time expended in each dwarf would be the same, increasing linearly the total time of the run.

Finally the computation of the value of the relic abundance is made with the aid of
\texttt{Micr\-OMEGAS}~\autocite{Barducci:2016pcb}.
To allow for the possibility of under-abundant DM candidate we penalize only relic
abundance values above the Planck experimental measured interval~\autocite{Aghanim:2018eyx},
points with a value of this observable conforming to or below the Planck value
are assigned a likelihood value of 
$1/(\sqrt{2\pi} \sigma)$ with $\sigma$ the Planck measured interval, this likelihood
value is added to the previously calculated one.

More technical details of the procedure described here can be found in the source codes of the modules used
to generate the results of this research, 
see~\autocite{colmo}.

\begin{figure}[!t]%
    \centering
\begin{tikzpicture}
\begin{axis}[
                ybar = 0.5,
                height=0.3\textheight,
                width=1.0\textwidth,
                ymin=0,
                xmin=0, xmax=47,
                enlarge x limits=0.05,
                xticklabels={$H^\pm_a$, $A_a$, $h_a$, $A$, $H$, $H_3$, $A_3$, $H^\pm_3$, $H^\pm$, $h$},
                bar width=32pt,
                bar shift=1.5,
                ylabel={Particle mass (TeV)},
                xlabel={Particle spectrum of BFP},
                nodes near coords,
                every node near coord/.append style={font=\footnotesize},
                axis lines*=left,
                y axis line style={opacity=0},
                yticklabels={\empty},
                ytick style={draw=none},
                xtick=data,
/pgf/number format/precision=6
            ]\addplot[fill=ballblue!20,draw=blue!60] coordinates { (0, 3.139745) (5, 3.139718) (10, 3.139717) (15, 0.5089) (20, 0.3836) (25, 0.3681) (30, 0.3441) (35, 0.3329) (40, 0.2789) (45, 0.125) };
            \end{axis}
\end{tikzpicture}
    \caption{Spectrum of masses of the scalar particles in the model evaluated at the BFP.}%
	\label{fig:bfp}%
\end{figure}
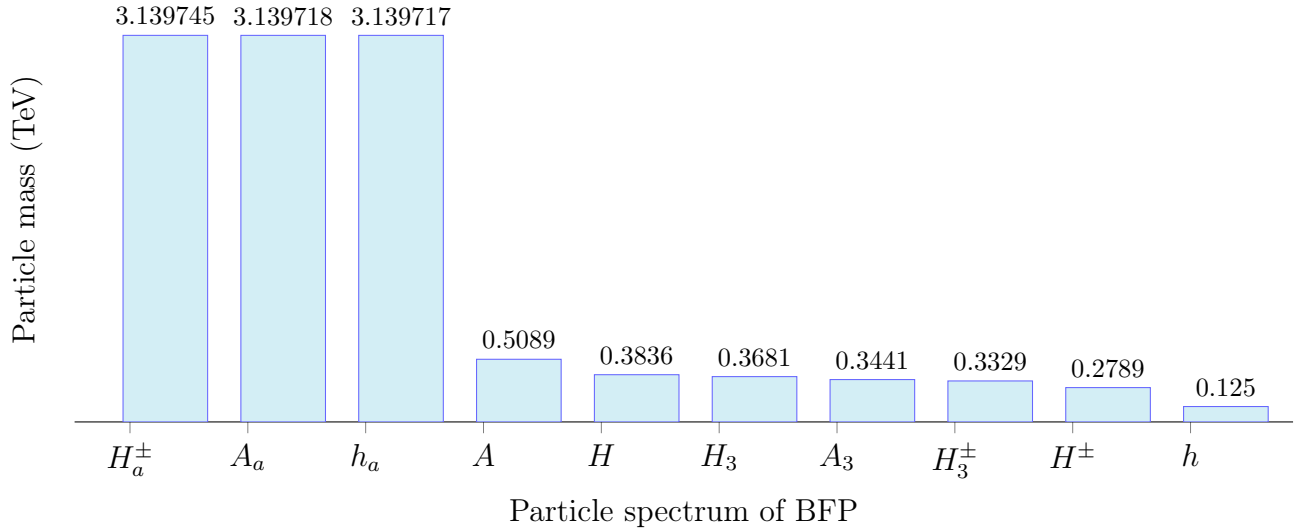

\section{Results}\label{results}

\begin{figure}[!ht]
	\centering
	\includegraphics[width=0.7\textwidth]{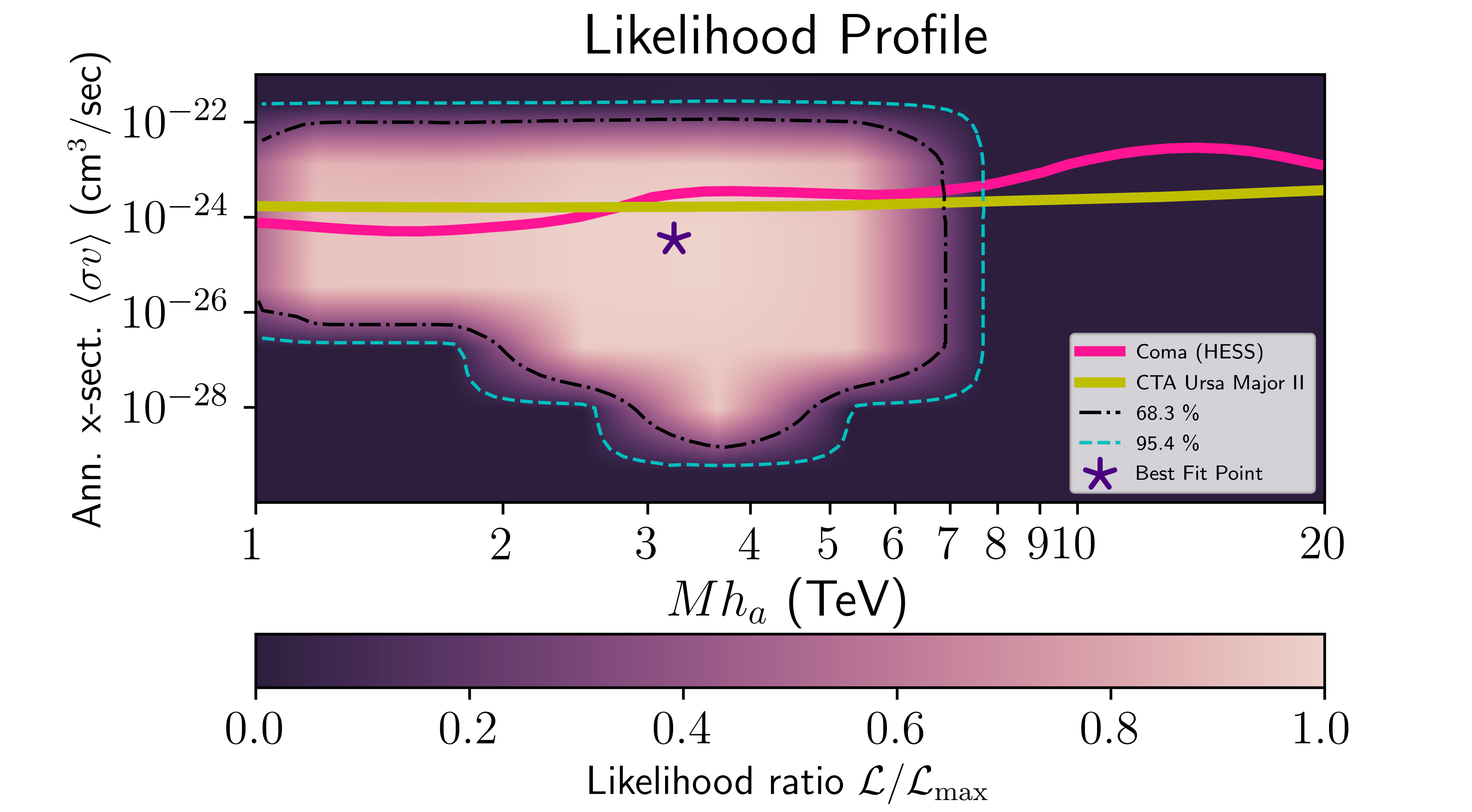}
	\includegraphics[width=0.7\textwidth]{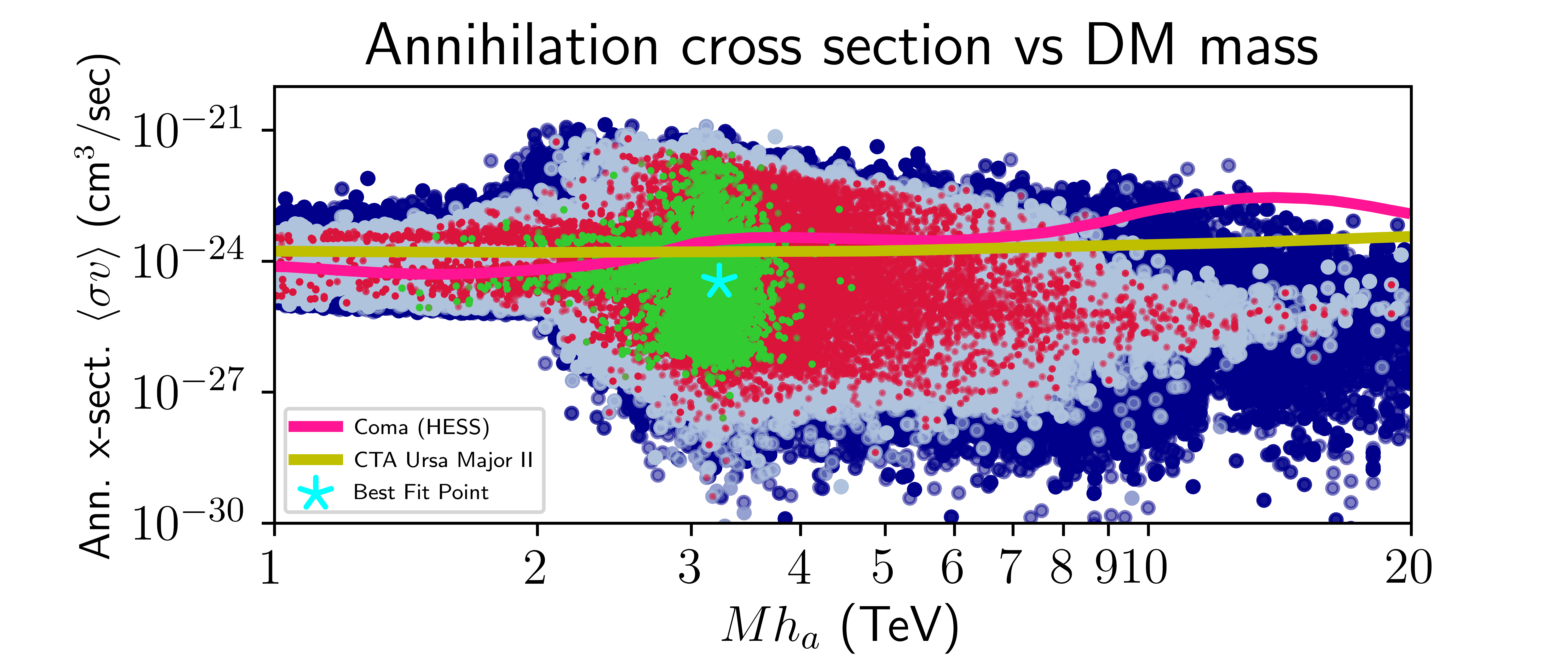}
	\includegraphics[width=0.7\textwidth]{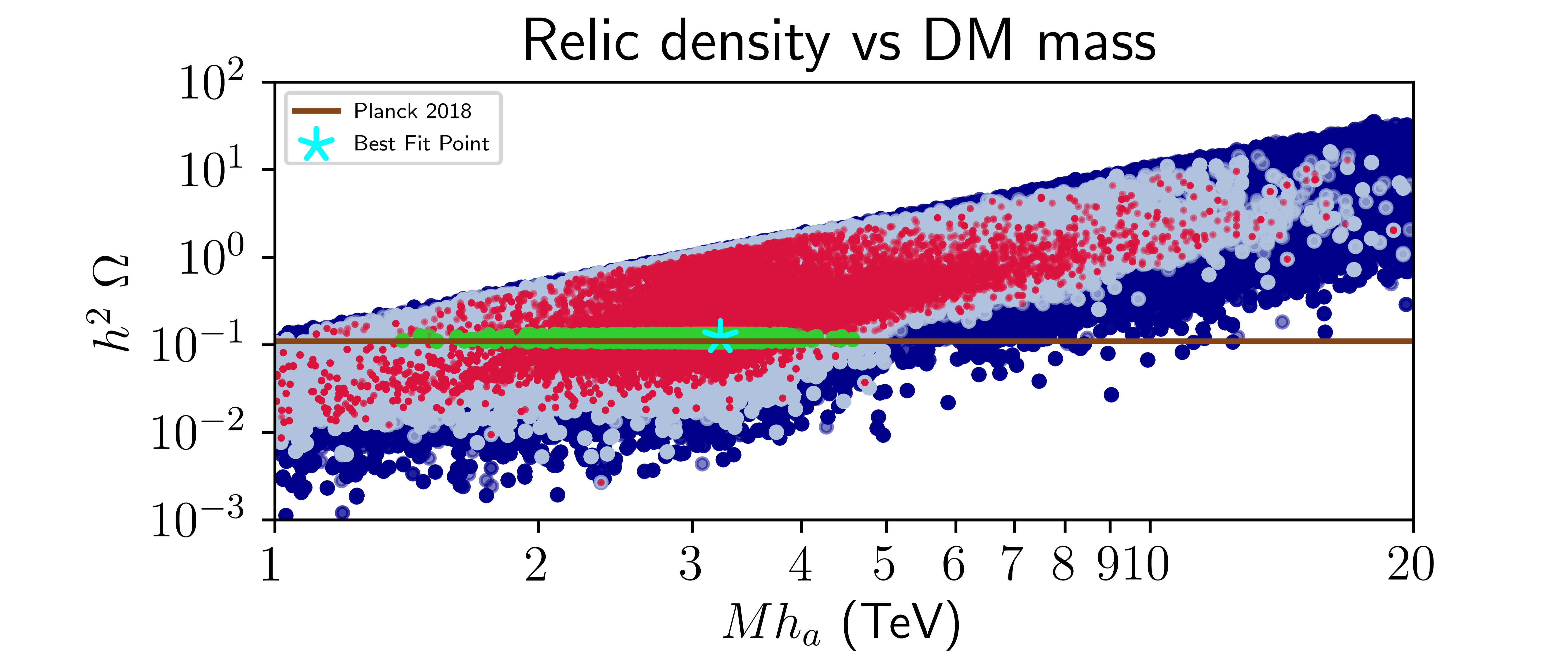}
	\caption{(Color online) Top panel: Normalized likelihood profile as a function of the zero-velocity averaged annihilation cross section and the DM candidate mass. Contours for a coverage probability of 68.3\% and 95.4\%
	for two degrees of freedom are shown as dot dashed and dashed respectively, and the BFP is marked as a star. For
	comparison we show exclusion curves from an analysis of an observation of Coma by HESS~\autocite{Abdalla:2018mve}
	and a simulation of a CTA observation of Ursa Major II~\autocite{Lefranc:2016dgx} (both model independent analysis).
	Middle panel: Scatter plot of the annihilation cross section as a function of the DM mass, navy blue points are
	consistent with unitarity and stability bounds but are excluded from experimental scalar searches; light blue
	points are consistent with all of these constraints while the red points in addition satisfy the decoupling limit
	for $h$; green points predict a relic abundance within the experimental Planck value~\autocite{Aghanim:2018eyx}.
	Bottom panel: Scatter plot of the predicted relic abundance as a function of the DM mass, the color code of the
	points is the same as the middle panel.
	}
	\label{fig:MvsS3}%
\end{figure}

We report here our results obtained from a final run with a population of 40 thousand points for the minimizer.
The spectrum of masses of the scalar sector at the best fit point is presented in figure \ref{fig:bfp}, the
DM sector masses are intentionally kept quasi-degenerate in order to facilitate the occurrence of the Sommerfeld
effect by allowing the production of on-shell DM intermediate states other than $h_a$ (by keeping the DM mass gaps small),
we find that the heavy (non DM) scalars lie with masses in between 278 and 509 GeV.

In figure \ref{fig:MvsS3} bottom panel, we present a scatter plot of points in the parameter space of the model
and their predictions of the value of the relic abundance as a function of the mass of the DM candidate, the points
are color coded according to weather or not they satisfy all physical constraints. With navy blue points excluded
from heavy scalar experimental searches, the rest of the points satisfy this constraint as well as unitarity and
stability and thus this plot makes it clear that only the region below the Planck value (the horizontal line)
represents physically acceptable points (assuming also the possibility of an under-abundant DM candidate). Therefore,
the region with DM masses above $\sim$5 TeV is clearly excluded. The difference between the light blue points and
the red ones is that for the latter the scalar $h$ is within the decoupling limit and thus has all the characteristics
of the SM Higgs particle. In addition, green points (a subset of the red ones) predict a relic abundance within
the experimental Planck interval, note that we found this kind of points within a region approximately from
$\sim$1.3 to $\sim$4.9 TeV masses.

In the middle panel of figure \ref{fig:MvsS3} we present a scatter plot of the annihilation cross section versus the
DM candidate mass with the same color code for the points as above. We include in the plot two exclusion curves
from independent analysis. The first, labeled Coma (HESS), is an analysis of the Coma Dwarf from observations 
with the HESS Imaging Atmospheric Cherenkov Telescopes~\autocite{Abdalla:2018mve} 
and represents a 95\% C.L. exclusion limit on the velocity
weighted cross section for DM self-annihilation into gamma ray lines ($\gamma\gamma$ lines). They also include
combined analysis with other dwarfs but we chose to present the comparison only with the limits from the observations
of Coma Berenices; note that this limits assume generic DM (or model independent). 
The second exclusion curve, labeled CTA Ursa Major II, is from a simulation of the CTA sensitivity to DM 
annihilation in the channel DM DM $\rightarrow W^+W^-$ assuming a 500 hour observation of the Ursa Major II
dwarf~\autocite{Lefranc:2016dgx}, also for generic dark matter; while in this reference there are also limits
from Coma Berenices (but only 50 hour observation simulation), we chose to compare with Ursa Major II 
because it has more stringent limits. We note from this scatter plot that there is a ``depression'' starting
around masses of 2 TeV where the cross section begins to take smaller values, this may be indicative
that the destructive interference in the enhancement factors found in the BFP (figure \ref{fig:somm1}) 
in between 2 and 3 TeV values of the DM mass is a generic feature of the parameter space and not only of the BFP.

Next, in the top panel of figure \ref{fig:MvsS3} we present the normalized likelihood profile as a function of the annihilation
cross section and the DM candidate mass, the best fit point of the analysis is marked as a star (in all panels);
we note how the region above a DM mass of 5 TeV is gradually disfavored mostly because it is already ruled out
by an overproduction of DM not consistent with observations. It is apparent from this plot that close to a third
of the viable region of plausible physical points lie above the two exclusion curves which are very similar
in the region of masses below 7 TeV, with the HESS limits slightly better below $\sim$3 TeV and the CTA
limits slightly better above this mass. Both exclusion curves come very close to disfavour the best fit point
of the model, probably with an improvement of the limits by an order of magnitude the BFP could be excluded.
Notice that we are assuming the possibility of under-abundant DM, had we imposed penalization of the likelihood
function also for points below the Planck interval by the assumption that this DM candidate comprises the totality
of the observed DM in the Universe, then the bright region of this plot would appear shrunken horizontally
reflecting the fact that green points in the other plots lie mostly around a mass of $\sim$3.1 TeV, as noted
before.

\begin{figure}[t]
	\centering
	\includegraphics[width=0.5\textwidth]{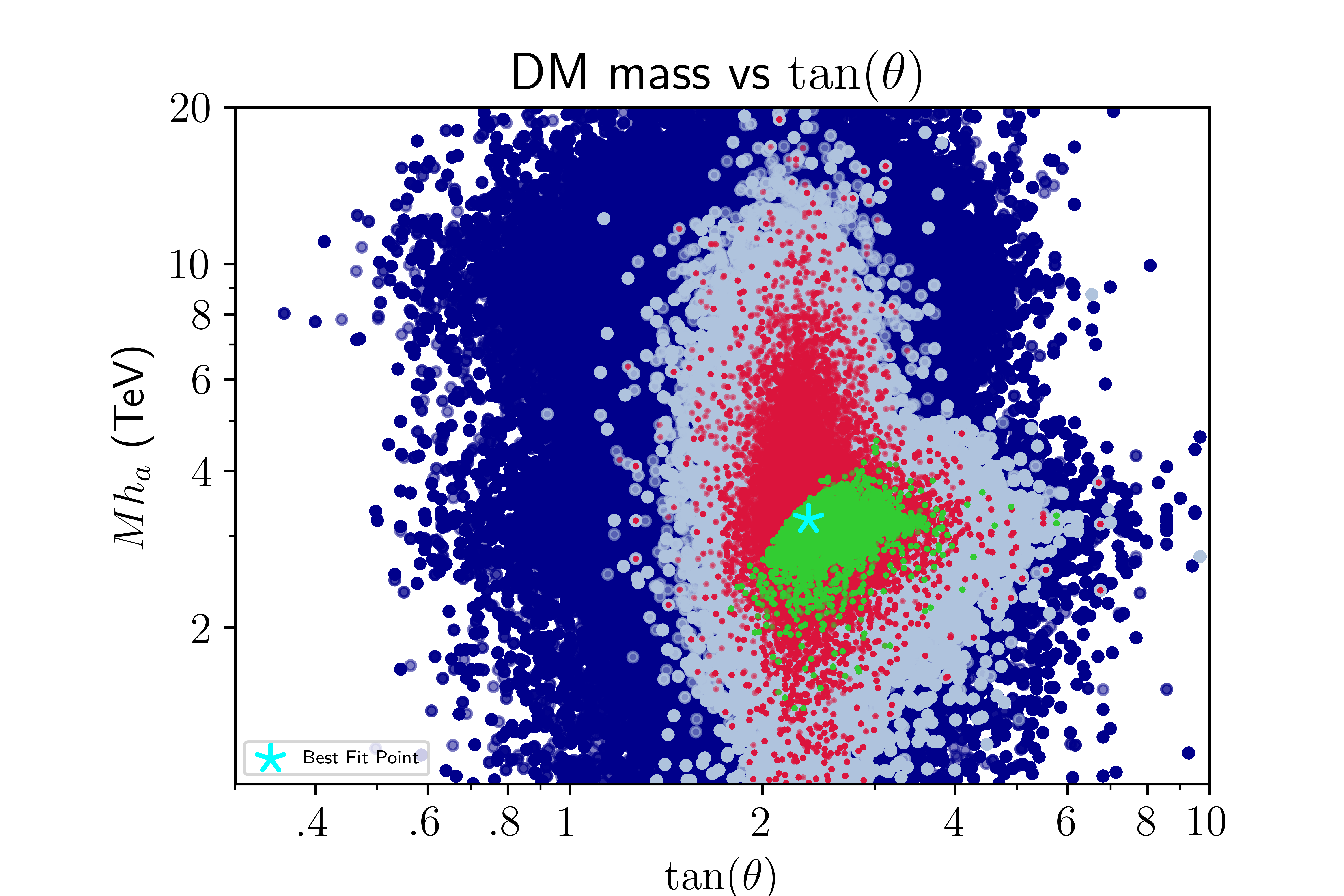}
	\caption{(Color online) Scatter plot of the DM mass as a function of $\tan\theta$, the color code 
	of the points is as in figure \ref{fig:MvsS3}.}
	\label{fig:tan}%
\end{figure}

In figure \ref{fig:tan} we show an scatter plot of the DM mass candidate as a function of $\tan\theta$
with the color coding the same as in the previous scatter plots. We see clearly the approximately formation
of subregions contained within regions of a lesser restraining level, thus points that satisfy the 
decoupling limit cover around two thirds of the region of points satisfying unitariry, stability
and Higgs searches, while points predicting a relic abundance value within the experimental interval
lie mostly between values of $\tan\theta$ equal 2 and 4, with the BFP attaining a value of 2.34.

\begin{figure}[t]%
    \centering
    \subfloat[]{{\includegraphics[width=0.5\textwidth]{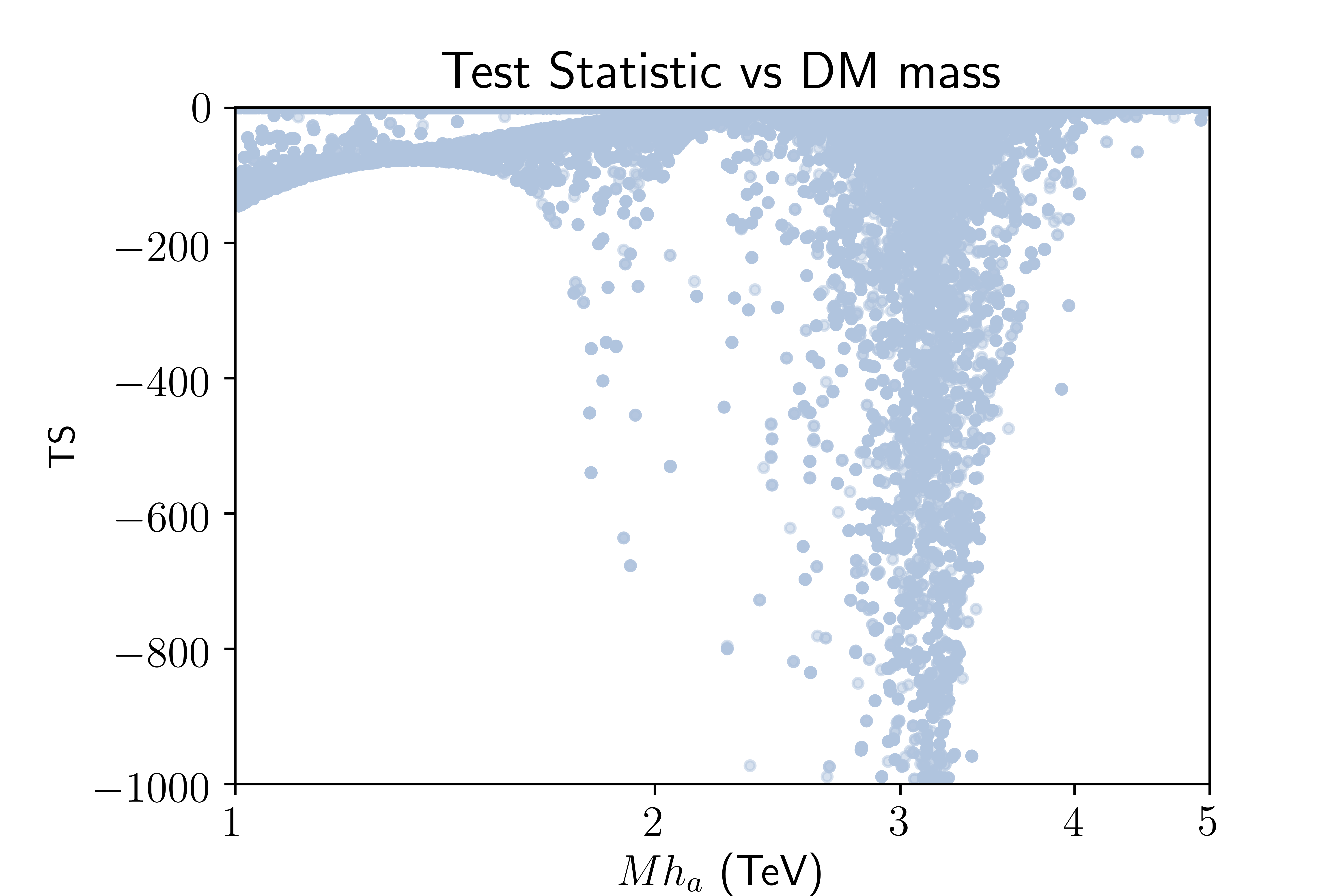} }}%
    \subfloat[]{{\includegraphics[width=0.5\textwidth]{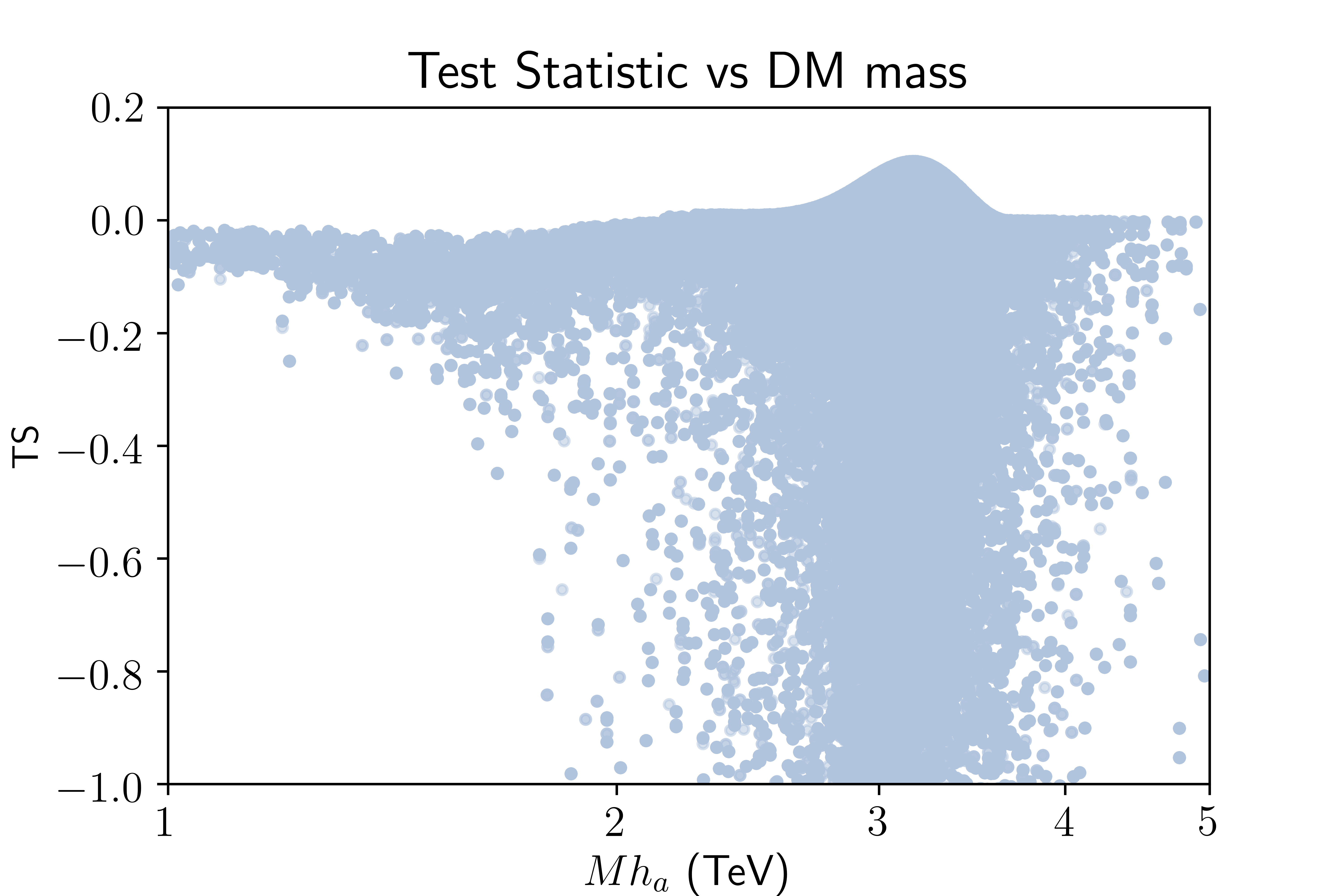} }}%
    \caption{(a) Scatter plot of the Test Statistic as a function of the DM mass, all the points shown satisfy
    unitarity, stability and experimental higgs searches constraints and in addition predict a relic abundance
    within the experimental Planck value or below it.
    (b) Same as (a) but shows only values above a TS > -1.0.}%
    \label{fig:TSvsM}%
\end{figure}

Since we are simulating a null-result experiment, the best indicator of the actual sensibility
of the experiment would be the computation of the Test Statistic to compare to what is expected
under the null hypothesis (no DM present), this technique is employed
in current ID experiments to construct exclusion limits from their non-observation of significant
gamma ray signals above the background. With this in mind, we present in figure \ref{fig:TSvsM}
(a) the scatter plot of the TS as a function of the DM mass in the DM mass range between 1 and 5 TeV,
this accounting for the fact that higher masses predict overproduction of DM in conflict with
the observed abundance as was discussed before. In fact, during the computational runs the calculation
of the CTA likelihood and the TS was not performed for points predicting a relic above the Planck
interval. Instead, these points were automatically assigned a ``bad'' value of the likelihood for purposes of 
optimization of the code. Figure \ref{fig:TSvsM} (b) shows the same plot but only for TS > -1.0, here
we note that the TS reaches an almost constant maximum value very close to zero for a given mass, with
a small bulb around a DM mass of $\sim$3.2 TeV, attaining a maximum TS close to 0.1.

\begin{figure}[t]
	\centering
	\includegraphics[width=0.8\textwidth]{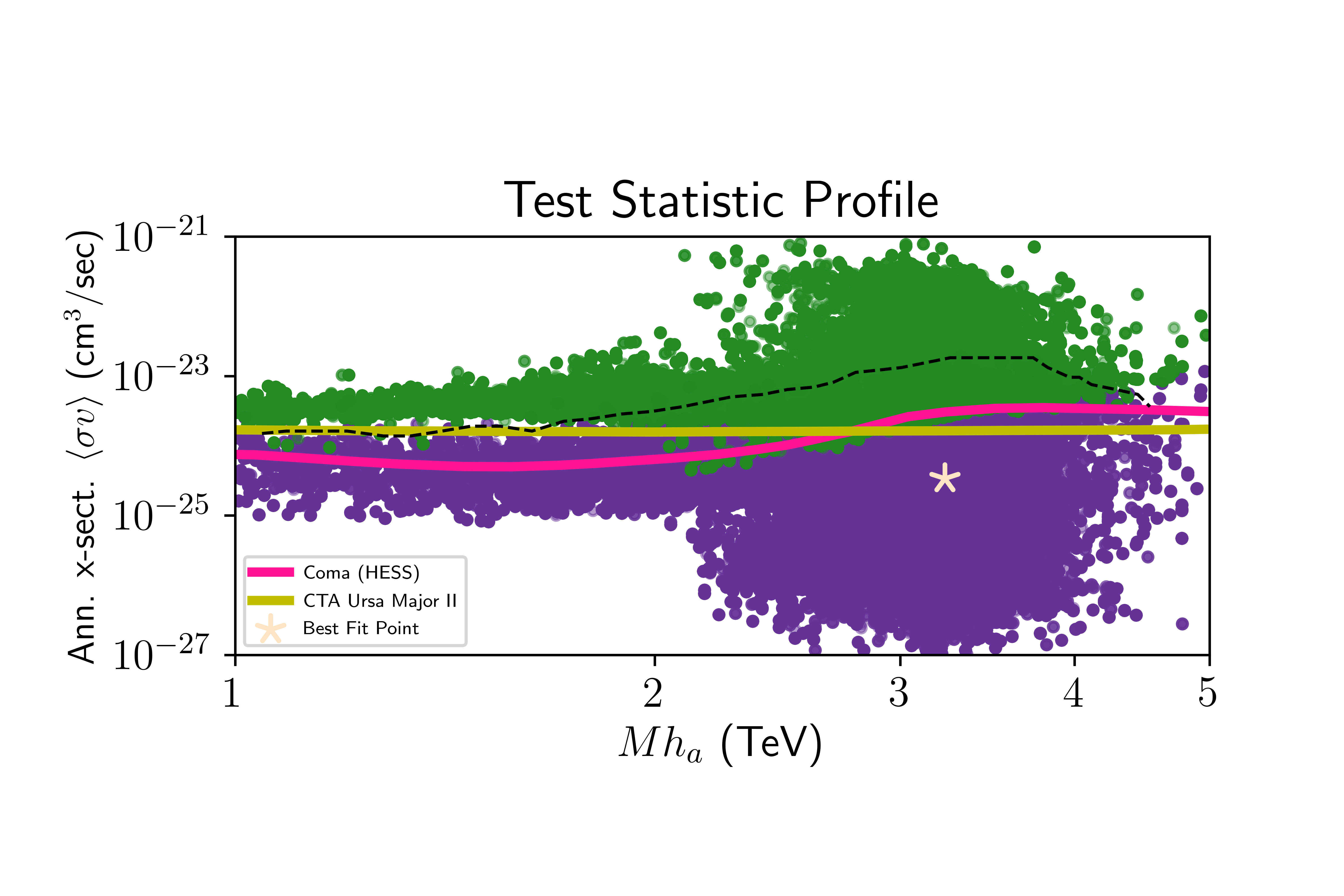}
	\caption{(Color online) Scatter plot of the annihilation cross section as a function of the DM mass with the
	points colored according to their value of the Test Statistic: purple points (mostly bottom region) have a
	TS less than 1 in absolute value, while green points (mostly top region) have a TS greater than 1 in absolute value. 
	The exclusion curves are as in figure \ref{fig:MvsS3} and the dashed black line shows the extend to which the purple
	and the green points overlap (for more details see main text).}
	\label{fig:ts}%
\end{figure}

For points in figure \ref{fig:TSvsM} lying in a (vertical) ray of constant DM mass, we consider the difference
$TS_{\textrm{max}} - TS \equiv \Delta TS$, since the Test Statistic approaches a $\chi^2$ function
in the large data sample limit, a value of $\Delta TS=1$ corresponds to a coverage probability
of 68.3\% for estimation of 1 parameter~\autocite{Tanabashi:2018oca}, 
which we take as the annihilation cross section.
Therefore, we interpret points in such a ray with $\Delta TS>1$ as excluded with a C.L. of 68\%.
In this manner we construct a ``Test Statistic profile'', shown in figure \ref{fig:ts} as a function of
the annihilation cross section and the DM mass. Since the value of $TS_{\textrm{max}}$ is very small,
for points such that $\Delta TS \gtrsim 1$
we approximate $\Delta TS \approx |TS|$ and the limit condition becomes $|TS|>1$ for them.
Following these arguments, we consider the points in the upper region of this figure excluded by
our analysis (at the corresponding C.L.) while the points in the bottom region are not. We can then 
interpret the boundary of the upper (green) region as an approximate exclusion curve, except that there is
an overlapping of the two regions, which we take as a consequence of the error bars of the analysis,
though a precise determination of these errors will not be pursued here.

\section{Conclusions}

We have presented in this letter an analysis of the Dark $S3$ model in the region of 
high dark matter candidate mass pursuing a determination of the prospects of indirect detection
specific to this model. In order to take into account most of the important theoretical 
predictions of the model we calculated the non-relativistic potentials which potentially
can cause enhancements on the value of the annihilation cross section by means of the 
Sommerfeld effect. We computed the enhancement factors and the annihilation matrices
leading to a proper determination of the cross sections and the differential gamma ray
flux from annihilation of DM pairs in the Coma Berenices dwarf galaxy. We determined
then the likelihood profile in parameter space from a simulation of observations of such
dwarf galaxy at the CTA assuming a null-result experiment using available public tools.
We conclude from our results that DM masses above 5 TeV are definitely exclude mainly
because of overproduction of DM not consistent with the observed value of the relic abundance.
For masses below 5 TeV the best fit point of the likelihood analysis suggest a DM value
of $\sim$ 3.14 TeV, somewhat higher than the value where the highest cross section enhancement
occurs indicating that other constraints such as unitarity and scalar searches are also
important for the precise determination of the best fit. Comparison of our results with 
independent analysis in the literature shows that current as well as estimated
exclusion limits are close to disfavour the model's BFP, at least under the conditions
of the present analysis. Perhaps the inclusion of a combined analysis taking into 
consideration several more dwarf galaxies will allow more stringent limits on the model.

\appendix

\section{Absorptive terms}\label{mats}

In this appendix we list the non-zero (symmetric) matrices of absorptive terms
for the calculation of the enhanced annihilation DM cross section
that complement equation (\ref{H3H3}).

Channel $\gamma \gamma$:
\begin{equation}\label{gammagamma}
\Gamma^{\gamma \gamma} = \frac{e^4}{4 \pi  M_{h_a}^2}
\times \mathrm{diag}(0,0,1)
\end{equation}

Channel $\gamma Z$:
\begin{equation}\label{gammaZ}
\Gamma^{\gamma Z} = \frac{e^4 \cot ^2(2 \theta_w)}
{4 \pi  M_{h_a}^2}
\times \mathrm{diag}(0,0,1)
\end{equation}

Channel $Z Z$:
\begin{equation}\label{ZZ}
\begin{array}{l@{}l}
\Gamma^{Z Z}_{11}      &{}=  \frac{1}{1024 \pi  M_{h_a}^2} 
(8 g_2^4 \sec ^4(\theta_w)+4 (-\cos (2 \theta) (\lambda_+ -\lambda_{14})+
\lambda_+ +\lambda_{14})^2) \\
\Gamma^{Z Z}_{12}      &{}=  \frac{1}{512 \pi  M_{h_a}^2}
 (4 g_2^4 \sec ^4(\theta_w)-4 \cos (2 \theta) \left((\lambda_{10}+\lambda_{11})^2-4 \lambda_{12}^2-\lambda_{14}^2\right)+\cos (4 \theta) (\lambda_+ -\lambda_{14}) (\lambda_- -\lambda_{14}) \\ 
 &{}+ 2 \lambda_{14} (\lambda_{10}+\lambda_{11})+3 (\lambda_{10}+\lambda_{11})^2-12 \lambda_{12}^2+3 \lambda_{14}^2) \\
\Gamma^{Z Z}_{13}      &{}=  \frac{1}{64 \sqrt{2} \pi  M_{h_a}^2}
 (g_2^4 \cos ^2(2 \theta_w) \sec ^4(\theta_w)+\lambda_{10} \sin ^2(\theta) 
 (-\cos (2 \theta) (\lambda_+ -\lambda_{14})+\lambda_+ +\lambda_{14})) \\
\Gamma^{Z Z}_{22}      &{}=  \frac{1}{1024 \pi  M_{h_a}^2}
 (8 g_2^4 \sec ^4(\theta_w)+4 (-\cos (2 \theta) (\lambda_- -\lambda_{14})+
 \lambda_- +\lambda_{14})^2) \\ 
\Gamma^{Z Z}_{23}      &{}=  \frac{1}{64 \sqrt{2} \pi  M_{h_a}^2}
 (g_2^4 \cos ^2(2 \theta_w) \sec ^4(\theta_w)+\lambda_{10} \sin ^2(\theta) (-\cos (2 \theta) 
 (\lambda_- -\lambda_{14})+\lambda_- +\lambda_{14})) \\
\Gamma^{Z Z}_{33}      &{}=  \frac{1}{64 \pi  M_{h_a}^2}
 (g_2^4 \cos ^4(2 \theta_w) \sec ^4(\theta_w)+2 \lambda_{10}^2 \sin ^4(\theta))
\end{array}
\end{equation}

Channel $W^+W^-$:
\begin{equation}\label{WW}
\begin{array}{l@{}l}
\Gamma^{W^+W^-}_{11}      &{}=  \frac{1}{512 \pi  M_{h_a}^2}
(4 g_2^4-4 \cos (2 \theta) (\lambda_+ -\lambda_{14}) (\lambda_+ +\lambda_{14})+\cos (4 \theta) (\lambda_+ -\lambda_{14})^2 \\ 
 &{}+  2 \lambda_{14} (\lambda_+)+3 (\lambda_+)^2+3 \lambda_{14}^2) \\
\Gamma^{W^+W^-}_{12}      &{}=  \frac{1}{512 \pi  M_{h_a}^2}
(4 g_2^4-4 \cos (2 \theta) \left((\lambda_{10}+\lambda_{11})^2-4 \lambda_{12}^2-\lambda_{14}^2\right)+\cos (4 \theta) (\lambda_+ -\lambda_{14}) (\lambda_- -\lambda_{14}) \\
&{}+2 \lambda_{14} (\lambda_{10}+\lambda_{11}) +3 (\lambda_{10}+\lambda_{11})^2-12 \lambda_{12}^2+3 \lambda_{14}^2) \\ 
\Gamma^{W^+W^-}_{13}      &{}=  \frac{1}{256 \sqrt{2} \pi  M_{h_a}^2}
(4 g_2^4+\lambda_{10} \cos (4 \theta) (\lambda_+ -\lambda_{14})+\lambda_{10} 
(3 (\lambda_+)+\lambda_{14})-4 \lambda_{10} \cos (2 \theta) \lambda_+) \\
\Gamma^{W^+W^-}_{22}      &{}=  \frac{1}{512 \pi  M_{h_a}^2} 
(4 g_2^4-4 \cos (2 \theta) (\lambda_- -\lambda_{14}) (\lambda_- +\lambda_{14})+\cos (4 \theta) 
(\lambda_- -\lambda_{14})^2+2 \lambda_{14} (\lambda_-) \\ 
&{}+3 (\lambda_-)^2+3 \lambda_{14}^2) \\ 
\Gamma^{W^+W^-}_{23}      &{}=  \frac{1}{256 \sqrt{2} \pi  M_{h_a}^2}
(4 g_2^4+\lambda_{10} \cos (4 \theta) (\lambda_--\lambda_{14})+\lambda_{10} (3 (\lambda_-)+\lambda_{14})-4 \lambda_{10} \cos (2 \theta) (\lambda_-)) \\
\Gamma^{W^+W^-}_{33}      &{}=  \frac{1}{256 \pi  M_{h_a}^2}
(4 g_2^4+\lambda_{10}^2 (\cos (4 \theta)-4 \cos (2 \theta))+3 \lambda_{10}^2) \\
\end{array}
\end{equation}

Channel $H H$:
\begin{equation}\label{HH}
\begin{array}{l@{}l}
\Gamma^{H H}_{11}      &{}=  \frac{1}{64 \pi  M_{h_a}^2}
(\left(\sin ^2(\alpha) (\lambda_+)+\lambda_{14} \cos ^2(\alpha)\right)^2) \\
\Gamma^{H H}_{12}      &{}=  \frac{1}{64 \pi  M_{h_a}^2}
(9 \left(\sin ^2(\alpha) (\lambda_-)+\lambda_{14} \cos ^2(\alpha)\right) \left(\sin ^2(\alpha) (\lambda_+)+\lambda_{14} \cos ^2(\alpha)\right)) \\ 
\Gamma^{H H}_{13}      &{}=  \frac{1}{32 \sqrt{2} \pi  M_{h_a}^2}
(9 \lambda_{10} \sin ^2(\alpha) \left(\sin ^2(\alpha) (\lambda_+)+\lambda_{14} \cos ^2(\alpha)\right)) \\
\Gamma^{H H}_{22}      &{}=  \frac{1}{64 \pi  M_{h_a}^2}
(\left(\sin ^2(\alpha) (\lambda_-)+\lambda_{14} \cos ^2(\alpha)\right)^2) \\ 
\Gamma^{H H}_{23}      &{}=  \frac{1}{32 \sqrt{2} \pi  M_{h_a}^2}
(9 \lambda_{10} \sin ^2(\alpha) \left(\sin ^2(\alpha) (\lambda_-)+\lambda_{14} \cos ^2(\alpha)\right)) \\
\Gamma^{H H}_{33}      &{}=  \frac{1}{32 \pi  M_{h_a}^2}
(\lambda_{10}^2 \sin ^4(\alpha)) \\
\end{array}
\end{equation}

Channel $h h$:
\begin{equation}\label{hh}
\begin{array}{l@{}l}
\Gamma^{h h}_{11}      &{}=  \frac{1}{64 \pi  M_{h_a}^2}
(\left(\cos ^2(\alpha) (\lambda_+)+\lambda_{14} \sin ^2(\alpha)\right)^2) \\
\Gamma^{h h}_{12}      &{}=  \frac{1}{64 \pi  M_{h_a}^2}
(9 \left(\cos ^2(\alpha) (\lambda_-)+\lambda_{14} \sin ^2(\alpha)\right) \left(\cos ^2(\alpha) (\lambda_+)+\lambda_{14} \sin ^2(\alpha)\right)) \\ 
\Gamma^{h h}_{13}      &{}=  \frac{1}{32 \sqrt{2} \pi  M_{h_a}^2}
(9 \lambda_{10} \cos ^2(\alpha) \left(\cos ^2(\alpha) (\lambda_+)+\lambda_{14} \sin ^2(\alpha)\right)) \\
\Gamma^{h h}_{22}      &{}=  \frac{1}{64 \pi  M_{h_a}^2} 
(\left(\cos ^2(\alpha) (\lambda_-)+\lambda_{14} \sin ^2(\alpha)\right)^2) \\ 
\Gamma^{h h}_{23}      &{}=  \frac{1}{32 \sqrt{2} \pi  M_{h_a}^2}
(9 \lambda_{10} \cos ^2(\alpha) \left(\cos ^2(\alpha) (\lambda_-)+\lambda_{14} \sin ^2(\alpha)\right)) \\
\Gamma^{h h}_{33}      &{}=  \frac{1}{32 \pi  M_{h_a}^2}
(\lambda_{10}^2 \cos ^4(\alpha)) \\
\end{array}
\end{equation}

Channel $H h$:
\begin{equation}\label{Hh}
\begin{array}{l@{}l}
\Gamma^{H h}_{11}      &{}=  \frac{1}{64 \pi  M_{h_a}^2}
(\sin ^2(\alpha) \cos ^2(\alpha) (\lambda_+-\lambda_{14})^2) \\
\Gamma^{H h}_{12}      &{}=  \frac{1}{64 \pi  M_{h_a}^2}
(9 \sin ^2(\alpha) \cos ^2(\alpha) (\lambda_--\lambda_{14}) (\lambda_+-\lambda_{14})) \\ 
\Gamma^{H h}_{13}      &{}=  \frac{1}{32 \sqrt{2} \pi  M_{h_a}^2}
(9 \lambda_{10} \sin ^2(\alpha) \cos ^2(\alpha) (\lambda_+-\lambda_{14})) \\
\Gamma^{H h}_{22}      &{}=  \frac{1}{64 \pi  M_{h_a}^2}
(\sin ^2(\alpha) \cos ^2(\alpha) (\lambda_--\lambda_{14})^2) \\ 
\Gamma^{H h}_{23}      &{}=  \frac{1}{32 \sqrt{2} \pi  M_{h_a}^2}
(9 \lambda_{10} \sin ^2(\alpha) \cos ^2(\alpha) (\lambda_--\lambda_{14})) \\
\Gamma^{H h}_{33}      &{}=  \frac{1}{32 \pi  M_{h_a}^2}
(\lambda_{10}^2 \sin ^2(\alpha) \cos ^2(\alpha)) \\
\end{array}
\end{equation}

\section*{Acknowledgements}

This research has made use of the CTA instrument response functions (version prod3b-v2) 
provided by the CTA Consortium and Observatory (for more details see~\autocite{CTA-performance}),
as well as the ctools package~\autocite{Knodlseder:2016nnv,ctools}, 
a community-developed analysis package for Imaging Air Cherenkov Telescope data. 
ctools is based on GammaLib, a community-developed toolbox for the scientific analysis of astronomical gamma-ray data.
Figure \ref{fig:coma3} was made with the aid of
SAOImage DS9~\autocite{2003ASPC..295..489J}.
C.E. acknowledges the support of Conacyt (M\'exico) C\'atedra 341. 
This research is partly supported by UNAM PAPIIT through Grant IN111518.

\printbibliography

\end{document}